# New Variable Stars Discovered by the APACHE Survey. II. Results After the Second Observing Season

**Mario Damasso**
INAF-Astrophysical Observatory of Torino, Via Osservatorio 20, I-10025 Pino Torinese, Italy;
Astronomical Observatory of the Autonomous Region of the Aosta Valley, fraz. Lignan 39, 11020 Nus (Aosta), Italy
E-mail address: damasso@oato.inaf.it ; m.damasso@gmail.com

**Lorenzo Gioannini**
Dept. of Physics, University of Trieste, Via Tiepolo 11, I-34143 Trieste, Italy

**A. Bernagozzi**
**E. Bertolini**
**P. Calcidese**
**A. Carbognani**
**D. Cenadelli**
Astronomical Observatory of the Autonomous Region of the Aosta Valley, fraz. Lignan 39, 11020 Nus (Aosta), Italy

**J. M. Christille**
Dept. of Physics, University of Perugia, Via A. Pascoli, 06123 Perugia, Italy;
Astronomical Observatory of the Autonomous Region of the Aosta Valley, fraz. Lignan 39, 11020 Nus (Aosta), Italy

**P. Giacobbe**
**L. Lanteri**
**M. G. Lattanzi**
**R. Smart**
**A. Sozzetti**
INAF-Astrophysical Observatory of Torino, Via Osservatorio 20, I-10025 Pino

## Abstract
Routinely operating since July 2012, the APACHE survey has celebrated its second birthday. While the main goal of the project is the detection of transiting planets around a large sample of bright, nearby M dwarfs in the northern hemisphere, the APACHE large photometric database for hundreds of different fields represents a relevant resource to search for and provide a first characterization of new variable stars. We celebrate here the conclusion of the second year of observations by reporting the discovery of 14 new variables.

# 1. Introduction

APACHE (A PAthway to the Characterization of Habitable Earths) is a ground-based photometric survey specifically designed to search for transiting planets orbiting bright, nearby early-to-mid M dwarfs (Sozzetti *et al.* 2013), and it is mainly sensitive to companions with orbital periods up to 5 days. The project is based at the Astronomical Observatory of the Autonomous Region of the Aosta Valley (OAVdA), located in the Western Italian Alps, and the scientific observations started in July 2012. The survey utilizes an array of five, automated 40-cm telescopes to monitor hundreds of M-dwarfs and, together with the search for transit-like signals in their light curves, we also look for new variables among the stars that fall in the fields of view of the telescopes, each centered on the target cool stars. We presented in Damasso *et al.* (2014) (hereafter Paper I) more than 80 new variable stars that we discovered after the first year of the survey. Here we announce a list of 14 new variables, not appearing in the AAVSO International Variable Star Index (VSX), that we have detected by the end of the second season of APACHE. The number of findings appears to be much less than those discussed in Paper I mainly because of some reasons. First, the number of new fields surveyed between 2013 and 2014 was not as large as that of the M dwarfs observed during the first season. Until the end of September 2013 we observed 257 different fields, while 190 between October 2013 and the end of September 2014. Of them, only 42 are newly observed fields. This large overlap between the targets observed over two consecutive seasons mainly reflects a key requirement of the survey. In fact, from a single observing site the detection of transiting planets with orbital periods up to 5 days naturally requires observations of the same targets which extend to more than one year. Moreover, weather conditions heavily influenced the observations during the second season, and the amount of data necessary for a good characterization of the variables was collected for a relatively low number of fields.

The time series of the APACHE differential magnitudes for the variables discussed here are available under request.

# 2. Instrumentation and methods

We refer the reader to Paper I, Christille *et al.* (2013) and Sozzetti *et al.* (2013) for a detailed description of the observation strategy, hardware and software systems which characterize the APACHE survey, that did not undergo any relevant change during the second season. For convenience, here we repeat only the main key parameters of the survey. Each APACHE telescope is characterized by a pixel scale of 1.5 arcsec/pixel and a field of view of 26' × 26'. While four telescopes observe in the Johnson-Cousin Ic filter, one instrument uses a V filter to observe the brightest M dwarfs, in order to increase the exposure times (that in in Ic band would be very short) and reduce the overheads. The APACHE observations are carried-out in focus, with exposure times, kept fixed during the sessions independently from the seeing (which typically ranges between 1.2 to 3 arcsec), that are in the range of 3 to 180 seconds and are optimized for the M dwarfs, which are the primary targets. The magnitudes of the target M dwarfs, over which the APACHE observing strategy is tailored, vary within 8–16.5 in the V band and 5.5–13 in the J band. We use a circular observing schedule, with each target being re-pointed typically ~20-25 minutes after the last observation. This cadence should be optimal for collecting enough data points which fall in the portion of the light curve showing a transit, if it is in progress, that usually is expected to last for 1 to 3 hours. Each time a target is pointed, three consecutive exposures are taken and usually the average value of the corresponding differential magnitudes is used for light curve analysis.

Light curves of M dwarfs and all the field stars are produced by the pipeline TEEPEE (see Paper I),

which we developed mostly in IDL[1] programming language specifically for APACHE. Basically, TEEPEE performs ensemble differential aperture photometry by automatically selecting *i*) the best set of comparison stars among the brightest in the field, i.e. those with light curves showing the lowest r.m.s, varying in number from field to field, and *ii*) the best aperture among twelve test aperture radii in the range 3.5 to 9 pixels. The selection is made only once for the primary target M dwarf, and the same set of comparison stars and aperture radius are applied to all the field stars.

As indicated in Paper I, also for the objects discussed here no spectroscopic observations and analysis could be performed to better characterize their astrophysical properties and validate our tentative variability classifications. Their classification is thus based to photometric data only. While the detection of new variables originated from the APACHE database, as done for Paper I we also examined the on-line light curve archive of the SuperWASP survey (http://exoplanetarchive.ipac.caltech.edu/), where data from the first public data release, collected between 2004 and 2008 for nearly 18 million targets, are stored. Where applicable, the SuperWASP data represent a very useful resource for a more accurate variability classification and determination of physical parameters, as for example the periodicity, of new variable stars. Each of the SuperWASP observatories[2] (one based in the island of La Palma, the other located at the site of the South African Astronomical Observatory) consists of eight wide-angle cameras (field of view 61 sq. degrees) that simultaneously monitor the sky for planetary transit events. The SuperWASP pixel scale is quite high (13.7 arcsec/pixel), and before using the data, we carefully checked if stellar companions were present so close to the variable of interest to fall within the SuperWASP aperture radius and contaminate the photometry.

Our results and the main information about the variables discussed in this work are listed in Table 1. In discussing some of the targets we used the information about the color excess *E(B-V)* integrated along the line of sight. We stress here that, without knowing the distance to the star, the color excess corresponding to the star location cannot be directly estimated, and we use the integrated E(B-V) along the line of sight only as a rough indication of how much the dust absorption can influence the correct interpretation of the photometric data. In all the cases the Galactic dust reddening for a line of sight was estimated through the maps provided by the NASA/ IPAC Infrared Science Archive (http://irsa.ipac.caltech.edu/applications/DUST/), using the determinations based on the work of Schlafly and Finkbeiner (2011).

As a final note, the finding charts of all the stars are publicly available through the Aladin Sky Atlas service (http://aladin.u-strasbg.fr/aladin.gml).

## 3. Results for individual variables

**UCAC4 837-000728**

The stellar map showing the location of this star is shown in the first panel of Fig. 1. We analyzed the light curve with the Lomb-Scargle (L-S) algorithm and detected a clear sinusoidal-like modulation with a dominant period of 0.185848 days. The L-S periodogram and the folded light curve are shown in Fig. 1. The periodicity and the amplitude of the light curve variations (~0.2 mag in *I* band) are compatible with those of a δ Scuti (DSCT) type, for which the highest frequency represents the principal pulsation mode, but the colors of the target are indicative of an early K rather than a A-F type star, if a main-sequence star is assumed. Nonetheless, the color excess E(B-V) integrated along the line of sight is

---

1   Registered trademark of Exelis Visual Information Solutions.
2   More details about SuperWASP can be found here http://www.superwasp.org/

~0.3 mag, and this suggests that the star could be significantly reddened due to interstellar dust.
We can confidently exclude that this object is an eclipsing binary because, by doubling the best period found, the typical shape of this type of variable does not appear and the original shape is not preserved in any form. Finally we note that two other peak frequencies appear in the periodogram, corresponding to periods of ~0.228271 and 0.156723 days in order of their significance. They could be interpreted as secondary pulsation frequencies.

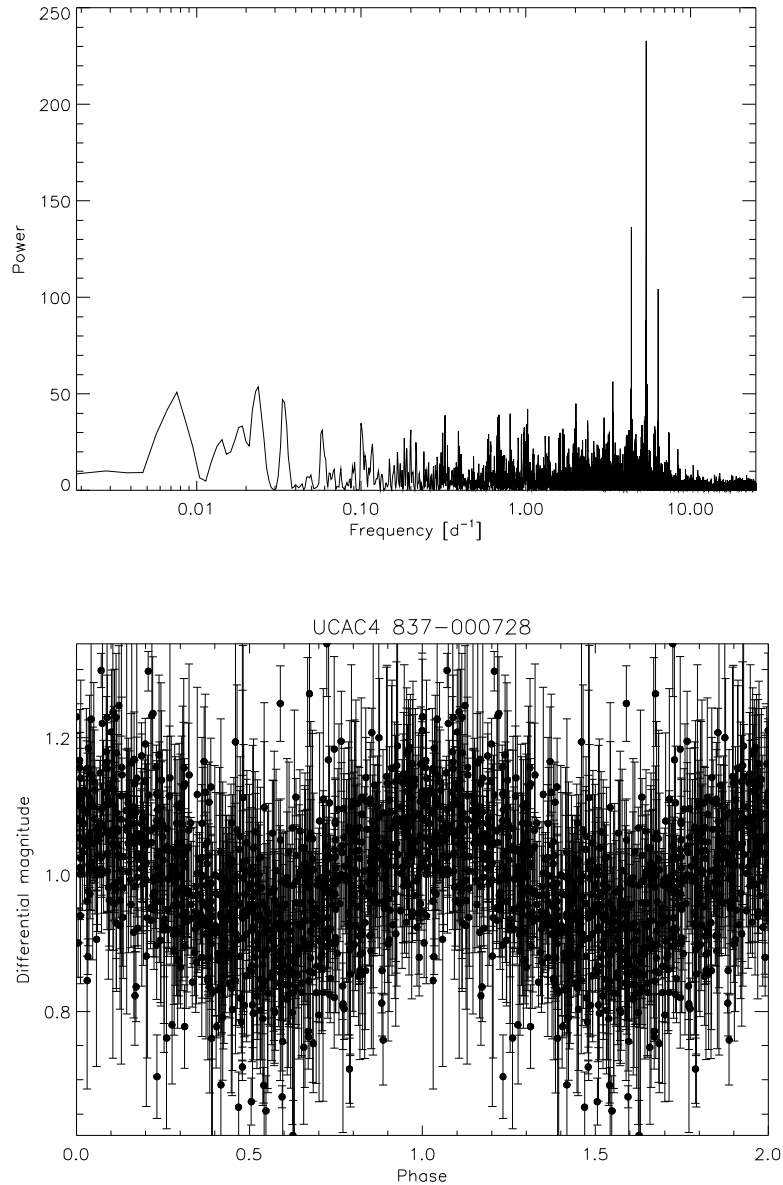

Fig. 1 – Star UCAC4 837-000728. (Upper plot) Lomb-Scargle periodogram of the APACHE photometric data. (Lower plot) APACHE light curve folded at the peak period P=0.185848 days.

**UCAC4 612-044588**

This variable appears to be a short-period (~21 hours) eclipsing binary showing a distorted light curve and a well-defined primary minimum (Fig. 2). Data from the APACHE survey, collected between January 8, 2013 and April 24, 2014, provide convincing evidence of the presence of the secondary minimum shifted of 0.5 in phase with respect to the primary. The star was also observed by the SuperWASP survey (more than 6600 data collected between September 23, 2004 and May 18, 2008), but the light curve (lower panel of Fig. 2) is characterized by a higher scatter than the APACHE time series, then making difficult the detection of the secondary minimum. The light curve distortion is also evident in the SuperWASP data and it is probably due to a direct physical interaction between the two components caused by their proximity.

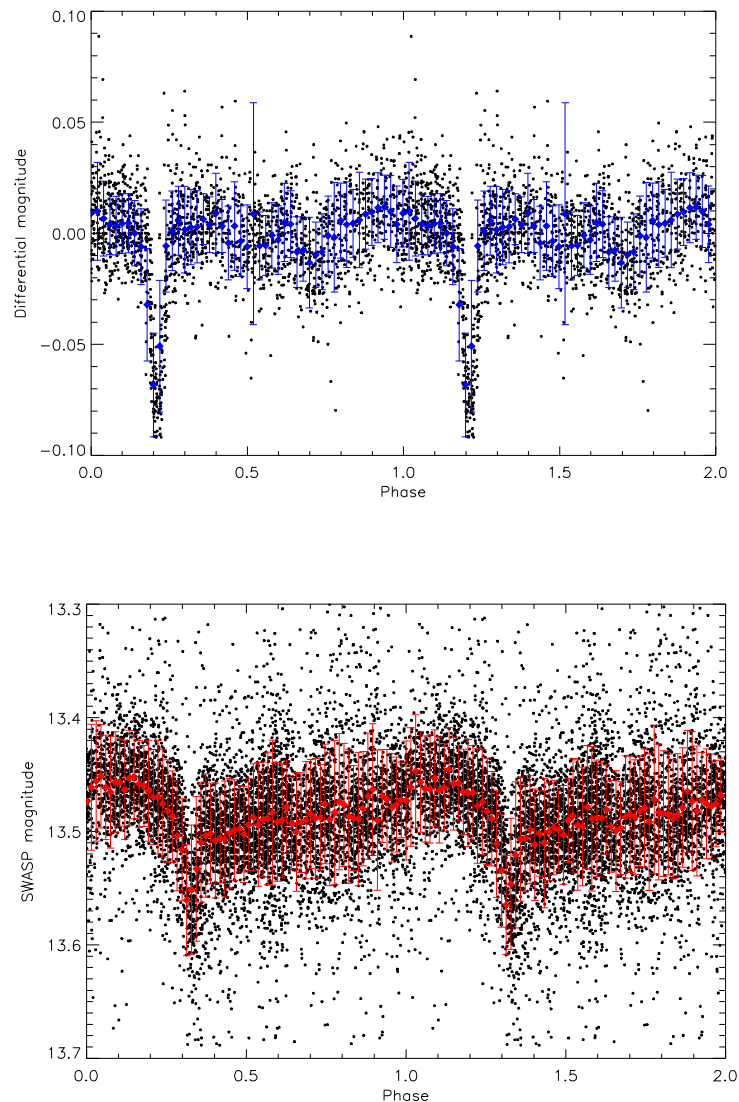

Fig. 2 - Eclipsing binary UCAC4 612-044588. (Upper plot) APACHE light curve *(I band)*, folded according to the best orbital period P=0.88965 days. (Lower plot) SuperWASP light curve, 3-sigma clipped and folded according to the same period.

# UCAC4 667-058562

The APACHE data for this target cover the quite limited timespan March 7- May 15, 2014. They show variations in the light curve that appear to have some periodicity that can be related to the rotation of the star (Fig. 3). In such a case, the flux modulation is produced by inhomogeneities in the stellar photosphere, as spots and active regions, that trace the mean rotation period of the star. By applying the Generalized Lomb-Scargle (GLS) algorithm (Zechmeister & Kürster 2009), we found a peak in the periodogram at $f_{max}$~0.0941 cycles/day (corresponding to P~10.62 days), and the data can be reliably fit with a sinusoid of semi-amplitude ~0.07 mag. Fitting the light curve of a star, which shows evidence of rotation, with a single sinusoid represents only a simplified model, because this describes the average structure of the photosphere over time and it does not take into account the life cycle of the active regions/spots and the changes occurring in their longitude distribution over the stellar disk. The nearly two-month APACHE observations suggest evidence of changes in the photospheric structures of the star over time with a time scale close to that of a single rotation cycle. This can be observed by comparing the different amplitudes of minima and maxima in the time series, reflected in the structure of the O-C residuals (upper panel of Fig. 3). To quantitatively assess the significance of the GLS peak frequency, we performed a bootstrap analysis (with re-sampling) of the APACHE data, from which it is possible to guess if the observed peak is real and at which level of confidence. It consists of two steps: *i)* randomly shuffling the magnitudes while keeping fixed the time stamps (allowing for multiple extractions of the same data point), and *ii)* performing a GLS analysis on the new dataset, with the same settings used to analyze the original data. If the observed signal is real, the shuffling should destroy without producing in the periodogram a higher peak corresponding to a different frequency *f*. On the contrary, if it is due simply to white noise, a more significant peak should appear, at any frequency, in several fake datasets. By repeating the bootstrap *N* times (*N*=10,000 in our case), we determined the total number of fake datasets for which the peak spectral power density, associated to any frequency, was higher than the spectral power density associated to $f_{max}$ in the case of the original dataset. It resulted that none of the fake datasets produced a spectral power density higher than the original, and for the P=10.62 days signal this corresponds to a False Alarm Probability (FAP) of $10^{-4}$, suggesting that it is not due to pure noise. FAP levels of 0.1% and 1%, respectively ten and one hundred times higher than that associated to $f_{max}$, correspond to a spectral power densities *p*=0.86 and *p*=0.83 , which are very close to power of the secondary peak in the original periodogram at *f*=0.075 cycles/day (P~13 days). This suggests that the secondary peak is not very significant but, due to our limited dataset, it could reflect the uncertainty we have about the stellar rotation period, or it could be related to the time scale over which the evolution of the active regions occurs.

The scenario emerging from the APACHE photometry is supported by the SuperWASP observations. This survey monitored the star, named 1SWASP-J135821.53+432058.4, between May 02, 2004 and April 18, 2008, collecting 13,768 points. The SuperWASP light curve, averaged in bins of 0.005 days, is shown in Fig. 4, both the folded data at P=10.62 days and the time series. The data, characterized by typical uncertainties higher than that of APACHE and covering a much longer timespan, show changes occurring in the light curve morphology over a time scale of the order of the rotation frequency. This strengthens the hypothesis of a rapid evolution of photospheric structures as spots and active regions. The GLS analysis applied to the SuperWASP data returned a peak at *f*~0.090 cycles/day (P~11 days), which confirms the result of the APACHE data analysis.

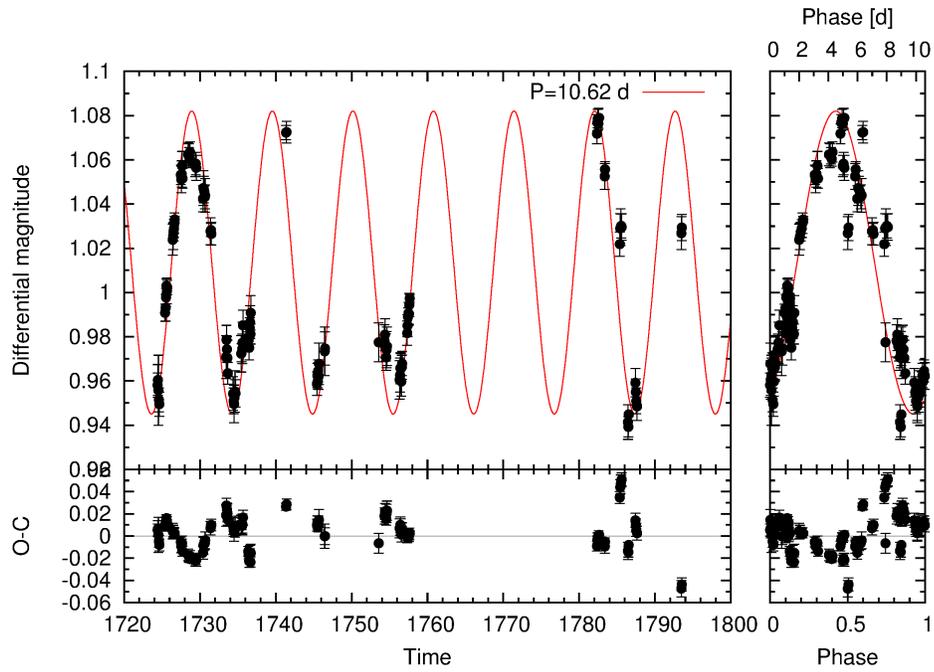
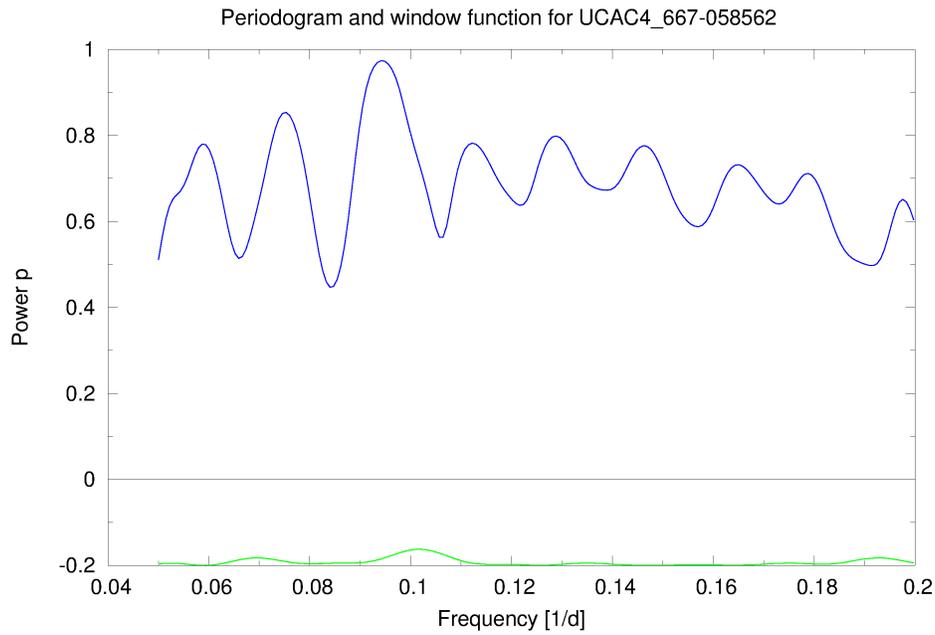

Fig. 3 – Star UCAC4 667-058562. (Upper panel) APACHE time series (upper left), and the folded light curve at P=10.62 days (upper right). Correspondingly, the residuals of a sinusoidal fit for both datasets are shown in the two lower plots. Time is provided as HJD-2,455,000. (Lower panel) GLS periodogram of the APACHE (blue curve). The best peak found is at f=0.0942 cycles/day. The green curve represents the window function of the APACHE observations.

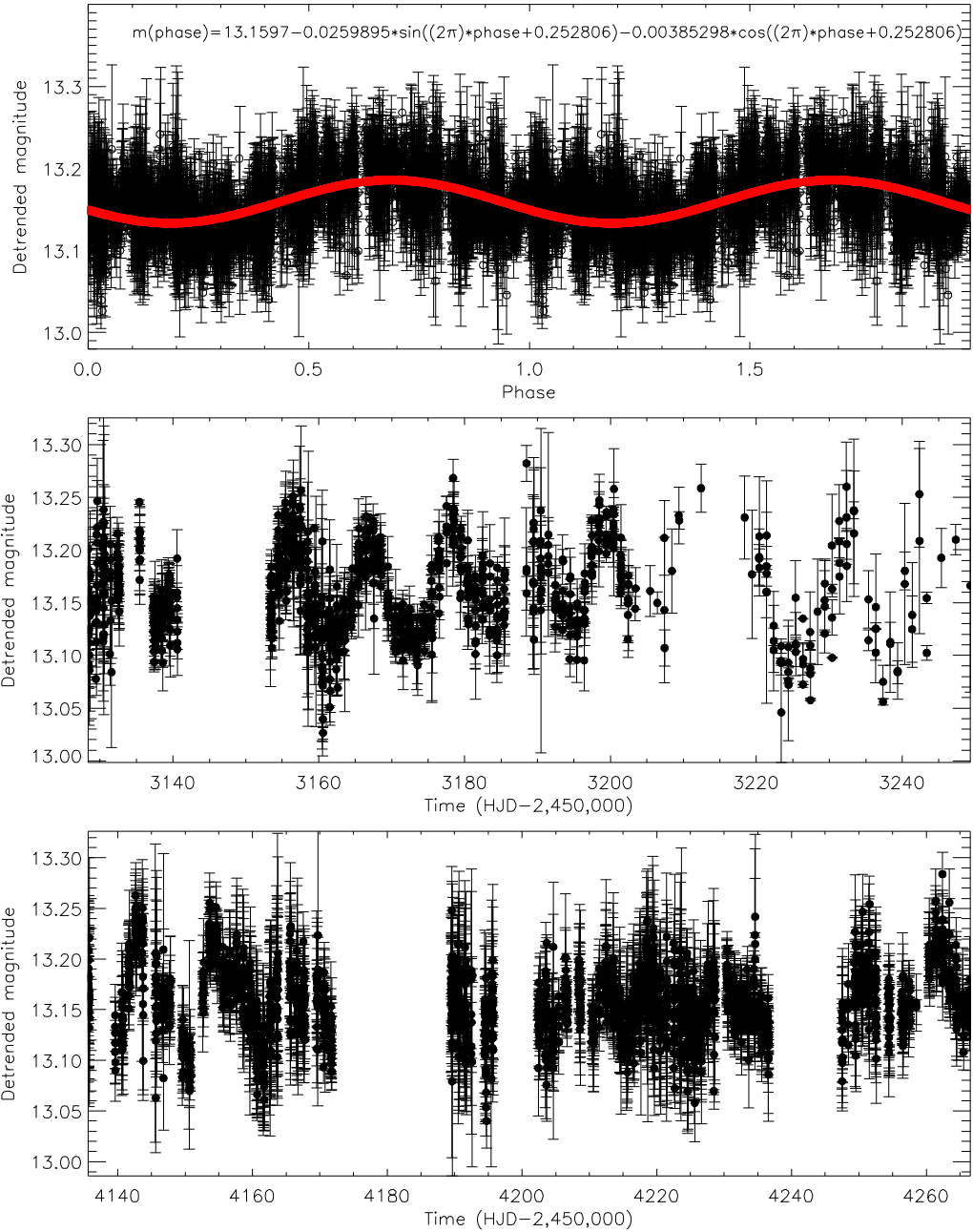

Fig. 4 – Light curve of the star UCAC4 667-058562 (1SWASP-J135821.53+432058.4) as observed by the SuperWASP survey. (Upper plot) Data folded according to the period P=10.62 days, with superposed the best fit function of the type A+B*sin(2π*$phase$+C)+D*cos(2π*$phase$+C). The curve consists of 3622 data points and was obtained by binning the original dataset in bins of 0.005 days and applying a 3-sigma clipping. (Middle and lower plots) Time series spanning almost 4 years of observations.

## UCAC4 854-011628

This object was observed by APACHE in *V* band between June 1 and October 21, 2014, collecting 1,405 useful points. Fig. 5 shows the light curve folded at P=0.30292 days, clearly indicating the nature of EW eclipsing binary for this target, with equal maxima and primary and secondary minima having a difference in their depths of ~0.03 mag. No data available from SuperWASP.

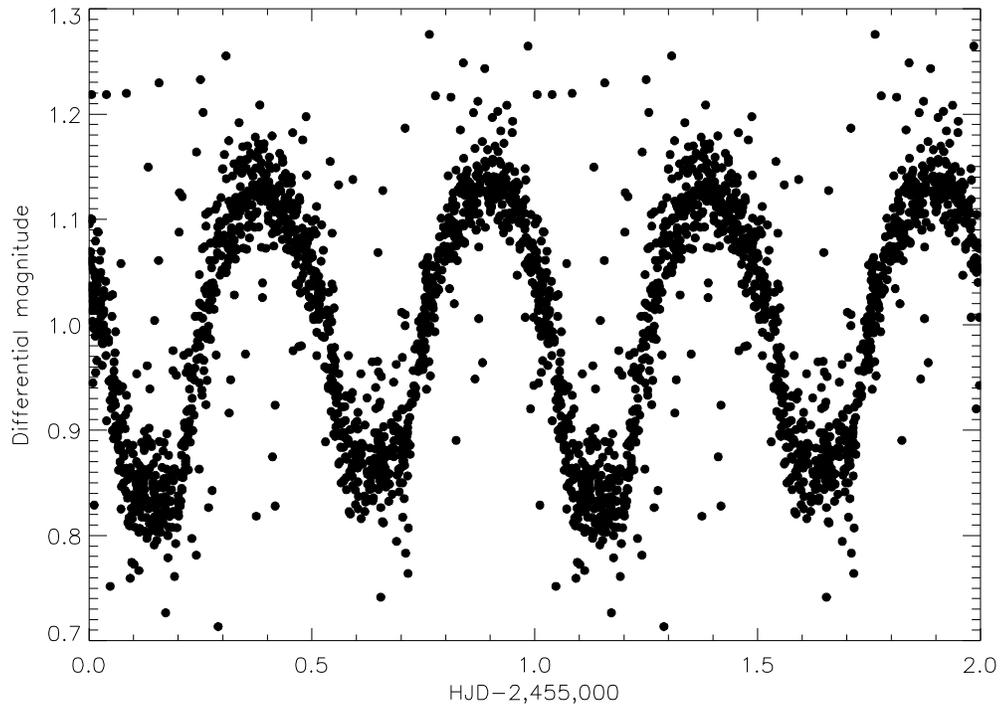

Fig. 5 – APACHE light curve in *V* band of the star UCAC4 854-011628, folded according to the period P=0.30292 days, clearly showing the existence of primary and secondary minima and a morphology typical of an EW eclipsing binary.

**UCAC4 609-091606**

We observed this target between July 12 and August 27, 2012, collecting 1,030 useful data points in *I* band. Fig. 6 shows the light curve folded according to the best, nonetheless tentative period found P=1.738 days, by refining the result of the analysis performed with the Box Least-Squares (BLS) algorithm (Kovacs *et al.* 2009). While the nature of eclipsing binary appears clear, from our data alone we cannot confirm the found periodicity, that can possibly be twice our estimate. Due to lack of data we cannot fully characterize the minimum we have observed but only provide a lower limit to its real depth, and we cannot explain what is the cause of the scattered portion of the light curve visible in the phase range [0.8, 1], which we could exclude to be due to instrumental noise and could actually be related to the existence of a secondary minimum. The solution corresponding to a period P=3.476 days is reliable, but the phase coverage of our data is not enough to prefer this hypothesis over the other. Unfortunately, no SuperWASP data are available to assess the real orbital period of this eclipsing binary.

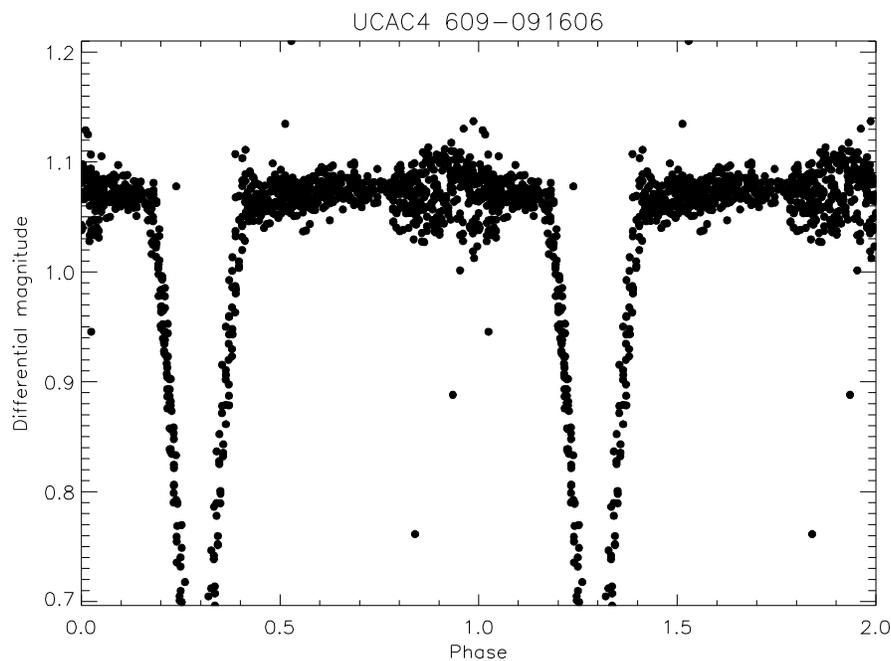

Fig. 6 – APACHE light curve of the star UCAC4 609-091606 folded according to the tentative period P=1.738 days.

**UCAC4 610-092815**

APACHE observed this object in *I* band between July 12, 2012, and August 14, 2014, for a total of 1,422 useful points. No SuperWASP observations are available. This variable appears to be an EW eclipsing binary, for which we estimate an orbital period P=0.42706 days. The folded APACHE light curve is shown in Fig. 7. Despite the light curve appears scattered due to the high magnitude of the star (V=15.741), because the observations with a 40-cm telescope were not optimized for this system, the two maxima show a different height, while the two minima appear to have the same depth (within the scattering level).

Due to a follow-up campaign which was focused on a different target than UCAC4 610-092815 but in the same field of view, we collected photometric data for this EW system in *V* band with the OAVdA 81-cm telescope. The observations were carried out during the nights of 13, 14 and 15 August, 2014, for a total of 566 useful data. The corresponding folded light curve is shown in the lower plot of Fig. 7. Due to the bigger aperture of the telescope, the data show much less scatter than those obtained in *I* band and confirm that the two maxima have not the same height, while the minima appear of the same depth also in *V*.

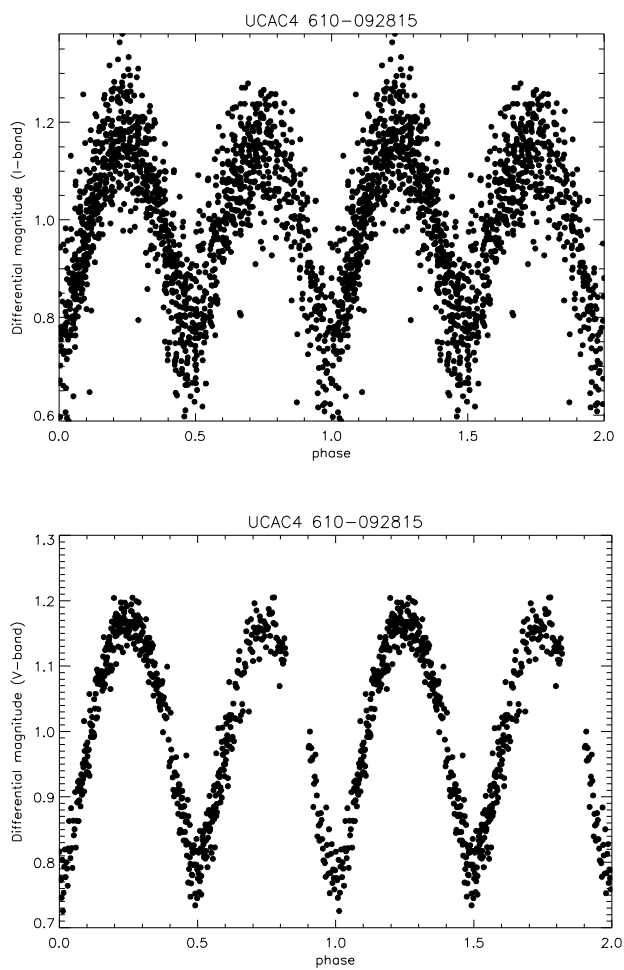

Fig. 7 – Star UCAC4 610-092815. (Upper plot) APACHE light curve folded according to the orbital period P=0.42706 days. (Lower plot) Data collected with the OAVdA 81-cm telescope in *V* band and folded according the same ephemeris.

**UCAC4 621-119831**

We observed this star in *V* band between June 21 and September 28, 2014, for a total of 1,011 useful points. The photometric time series is shown in Fig. 8, from which it appears that the variable changes its magnitude in an irregular way, at least from our dataset. We tentatively classify this object as L variable. The flux of the star appears to be red-dominated (B-V=1.757, V-Ks=4.759, J-H=0.909, H-Ks=0.268), with a reddening weighing on at level of a tenth (or less) of magnitude. The color indexes are compatible with those of an early M giant (Zombeck, 1990), and appear not compatible with those of a red dwarf. Without a reconnaissance spectrum available, our analysis is necessarily based on the expected intrinsic colors. Taking as a reference the intrinsic colors derived by Pecaut & Mamajeck (2013) for main sequence stars (see Tab. 4 therein), it can be seen for example that the combination of the *V-J*, *V-K* and *J-H* color indexes is not representative of an M dwarf, also taking into account their uncertainties, which are of the order of few hundredths of magnitude (Table 1).

Looking at the finding chart (Fig. 8), we note that UCAC4 621-119831 is very close to the much fainter star UCAC4 621-119835, which is separated by ~5'' from the target, or ~3.5 pixels in one APACHE image. No measurements of the *V* magnitude are available in the VizieR archive for this visual companion, which is ~5 magnitudes fainter in *J band* (2MASS photometry). Our data are not corrected for blending, but we believe that, considered the large difference in luminosity between the pair of objects, the level of contamination should be very negligible in this case.

No SuperWASP data are available for this object.

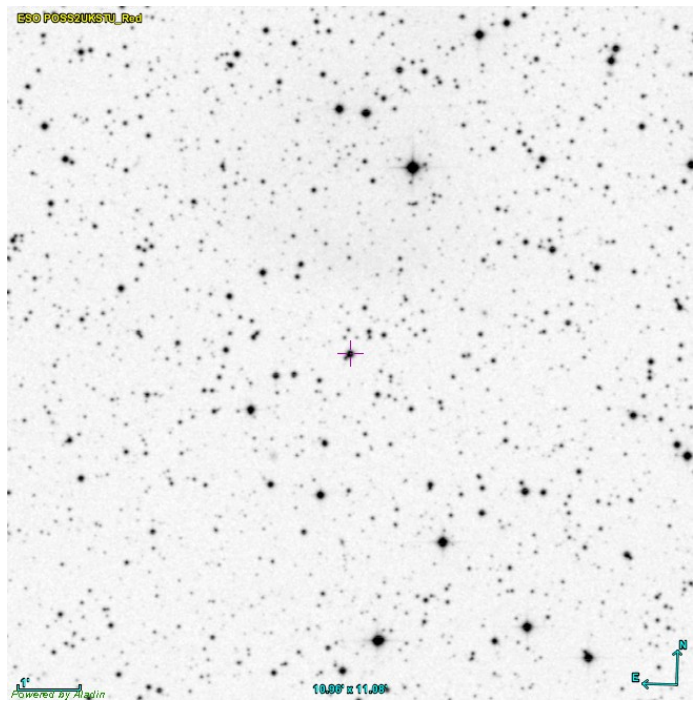

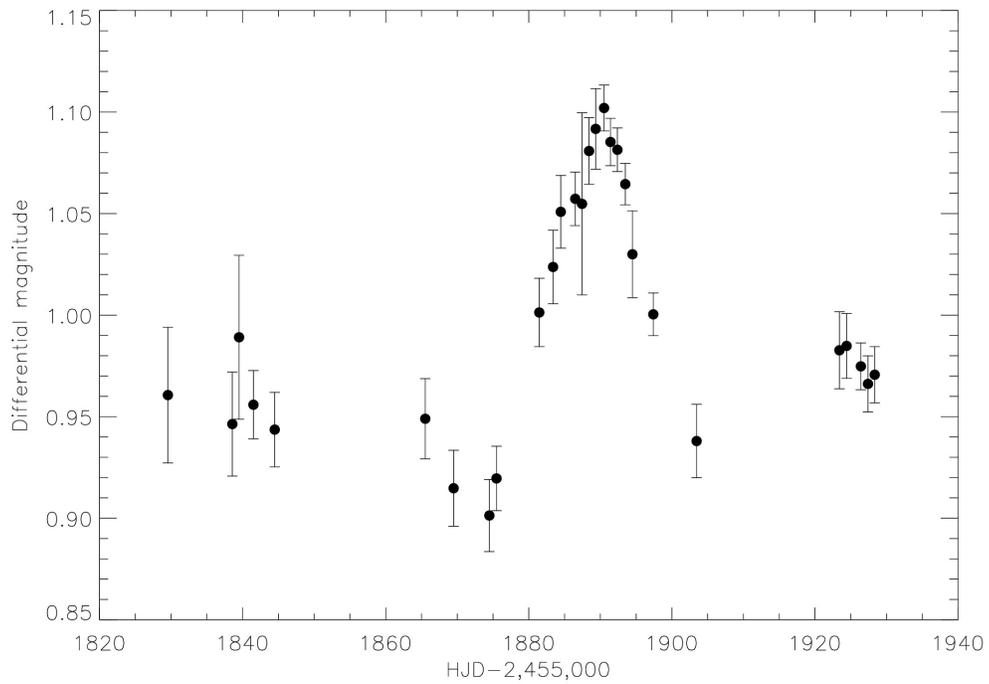

Fig. 8 - (Upper figure) Sky chart indicating the object UCAC4 621-119831 (red cross in the middle of the image). The image was downloaded from the Aladin sky atlas service and has linear dimensions 10.96'x11.08'. The image scale (1 arcsec) is indicated in the lower left corner. (Lower plot) APACHE photometric time series. Each point represents the average of the data collected during a single night of observation, and the error bars are the r.m.s. of the data of the corresponding night.

**UCAC4 620-119316**

This star shares the same field of UCAC4 621-119831 in the APACHE scientific frames, then it was observed in the same time span, collecting 954 useful points. The APACHE time series is shown in Fig. 9. The star appears as a single star, with no risk of blending with another object, and its flux is red-dominated, with color indexes similar to those of UCAC4 621-119831 (B-V=1.842, V-Ks=4.723, J-H=0.94, H-Ks=0.246). For the same reasons discussed for the previous variable, the color indexes do not appear compatible with the intrinsic values expected for a M dwarf star, suggesting this also should be an M giant. The light curve is indicative of an L irregular variable.

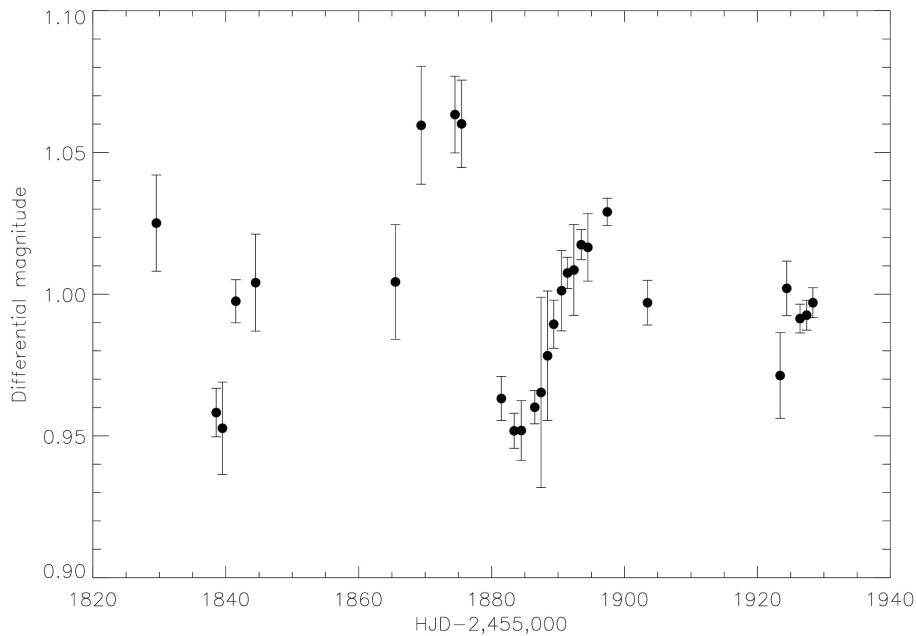

Fig. 9 - APACHE photometric time series of the star UCAC4 620-119316. Each point represents the average of the data collected during a single night of observation, and the error bars are the r.m.s. of the data of the corresponding night.

**UCAC4 620-119722**

This star was observed by APACHE in *V* band between June 21 and September 28, 2014, for a total of 974 useful points. We found photometric data also in the SuperWASP archive, covering the period between June 23, and October 26, 2007, for a total of 2198 points. The Lomb-Scargle algorithm results in almost the same periodogram for both the datasets, with a best period of 0.2778 days. The folded light curves (see Fig. 10 and 11) appear to have a quite sinusoidal shape, with data from APACHE much less scattered than those of SuperWASP. We note that for the SuperWASP dataset the second more relevant peak in the L-S periodogram (at 0.3850 days) is the third significant peak in the APACHE periodogram, while the third SuperWASP peak (at 0.2173 days) is the second for relevance for the APACHE dataset.

By looking at the index colors listed in Table 1, and taking into account the non-negligible, integrated reddening along the line of sight, this variable could be tentatively classified as an early-mid F-type main sequence star (Zombeck, 1990; Pecaut & Mamajeck, 2013). The spectral type, the periodicity and the amplitude of the light curve variations are compatible with those of a $\delta$ Scuti, which we then propose as a reliable classification.

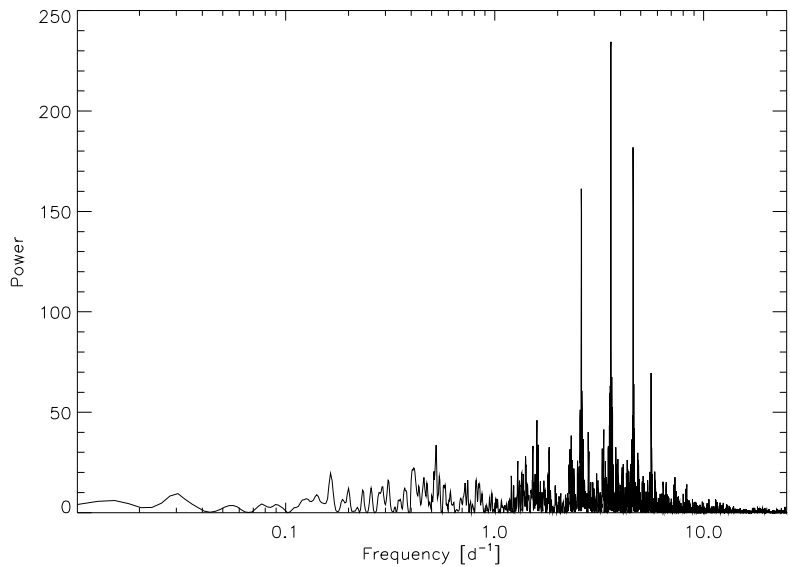

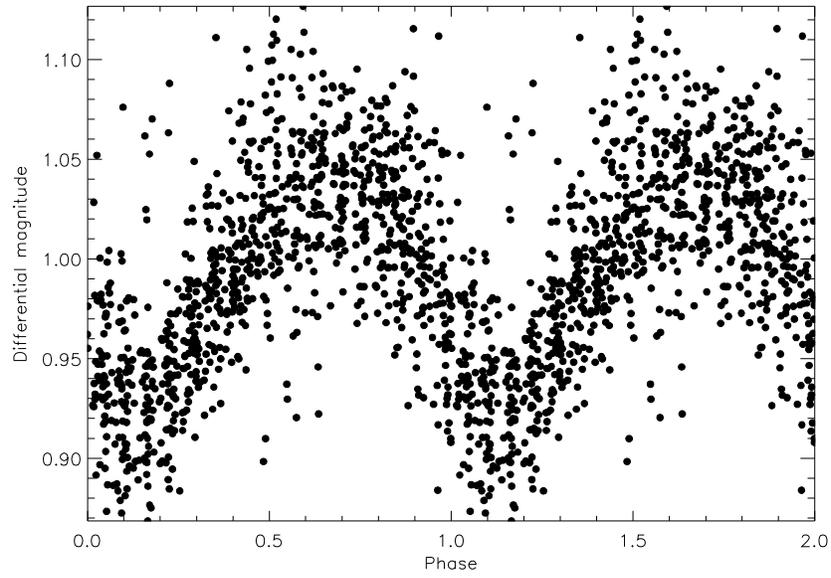

Fig. 10 - Star UCAC4 620-119722. (Upper plot) Lomb-Scargle periodogram for the APACHE data. (Lower plot) APACHE data folded according to the period 0.2778 days.

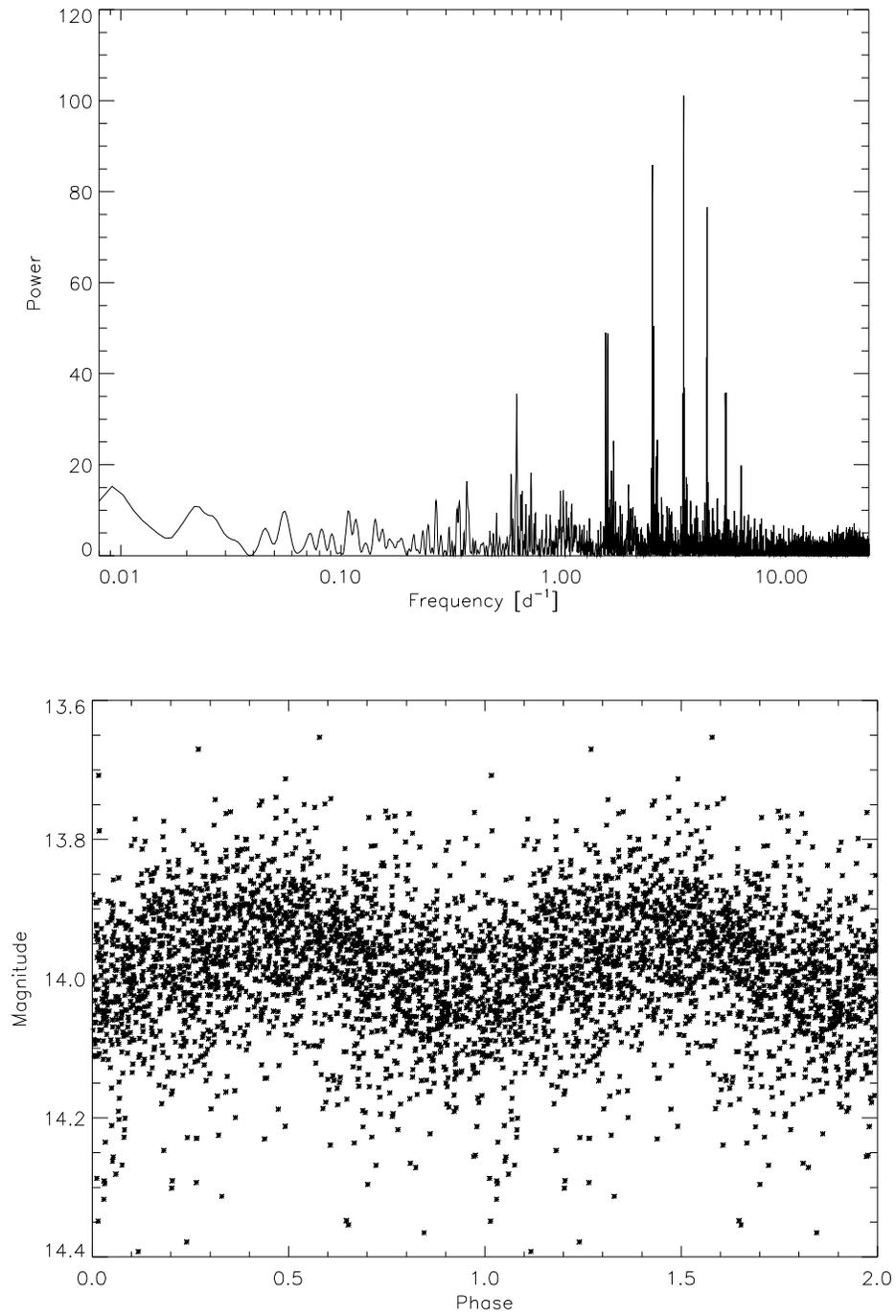

Fig. 11 - (Upper plot) Lomb-Scargle periodogram for the SuperWASP data of the star UCAC4 620-119722. The best period coincides with that found in the APACHE data. (Lower plot) SuperWASP light curve folded according to the best period 0.2778 days.

## UCAC4 673-106048

We observed this star between August 28, 2013, and September 28, 2014, for a total of 797 useful measurements in *I* band. The APACHE photometry is shown in Fig. 12, revealing a clear variation in the star luminosity with no periodic modulation. For the same reasons previously discussed, this star is characterized by color indexes that appear to be not representative of a dwarf star, but rather they are more typical of a M giant (B-V=1.92, V-K=7.281, J-H=0.996, H-Ks=0.423). In particular, while the V-K index is compatible with that of a M6V star, the observed V-J index appears ~0.5 mag lower than expected value (and this difference is expected to be even higher, if we consider the reddening correction that lowers the V-J index), and J-H is ~0.35 mag higher than the tabulated value. We therefore classify it as an LB-type (slow irregular of late spectral type) pulsating variable.

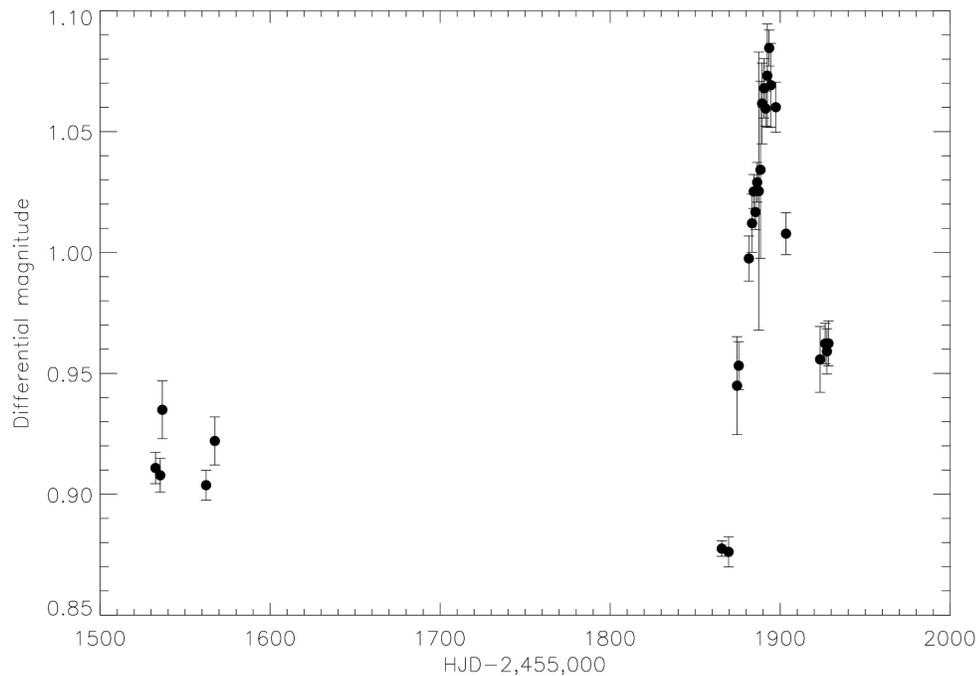

Fig. 12 - APACHE photometric time series of the star UCAC4 673-106048. Each point represents the average of the data collected during a single night of observation, and the error bars are the r.m.s. of the data of the corresponding night.

## UCAC4 858-013784

APACHE observed this star between April 3 and October 2, 2013, with 1,574 useful data in *I* band. Fig. 13 shows the Lomb-Scargle periodogram of our measurements characterized by a peak at 0.07077 days, and the light curve folded at this period, characterized by amplitude of ~0.02 mag. The color indexes (see Table 1, and also J-H=0.177, H-Ks=0.059), taking into account the magnitude of the integrated reddening term E(B-V), appear to be compatible with those of a late F-type main sequence star (e.g. Pecaut & Mamajeck, 2013). Together with the light curve characteristics, this supports the classification of the variable as a δ Scuti star. No SuperWASP data are available.

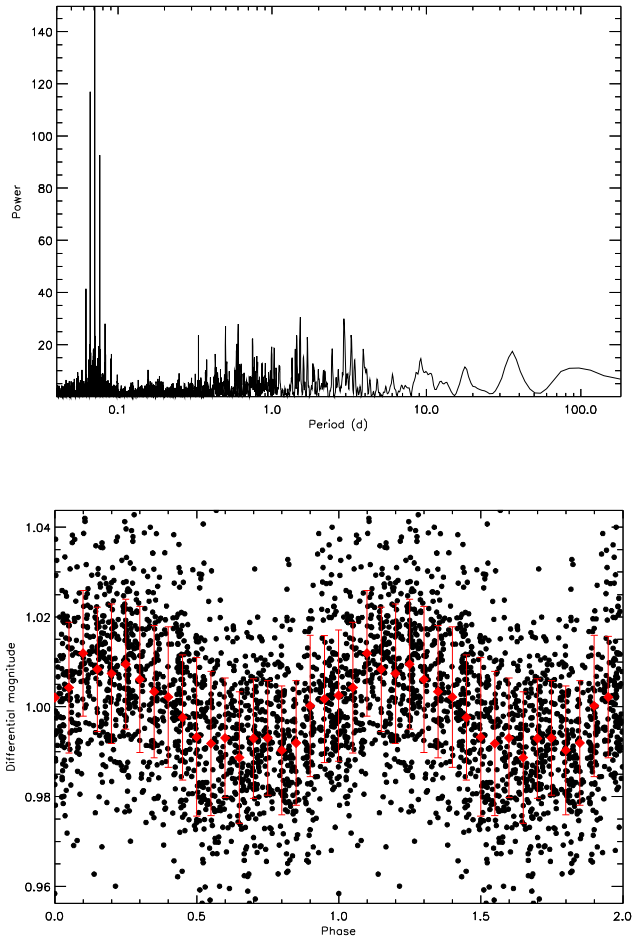

Fig. 13 - Star UCAC4 858-013784. (Upper plot) Lomb-Scargle periodogram for the APACHE data. (Lower plot) APACHE data (complete dataset in black and average values within bins of 0.05 in red) folded according to the period 0.07077 days.

**UCAC4 849-017521**

This object was observed in *I* band from 2013, March 16, to 2014, June 2, for a total of 2833 measurements. We applied the BLS algorithm to the light curve obtained by normalizing the measurements of each single night to their median value. This variable resulted to be an eclipsing binary of EW type, with an orbital period P=0.5988±0.0001 days. We show in Fig. 14 the light curve binned at 0.005 days and folded according to the period found, which clearly reveals the presence of a primary and a secondary minimum with different depths, while the two maxima appear equal within the scatter of our data. No SuperWASP data are available for this variable

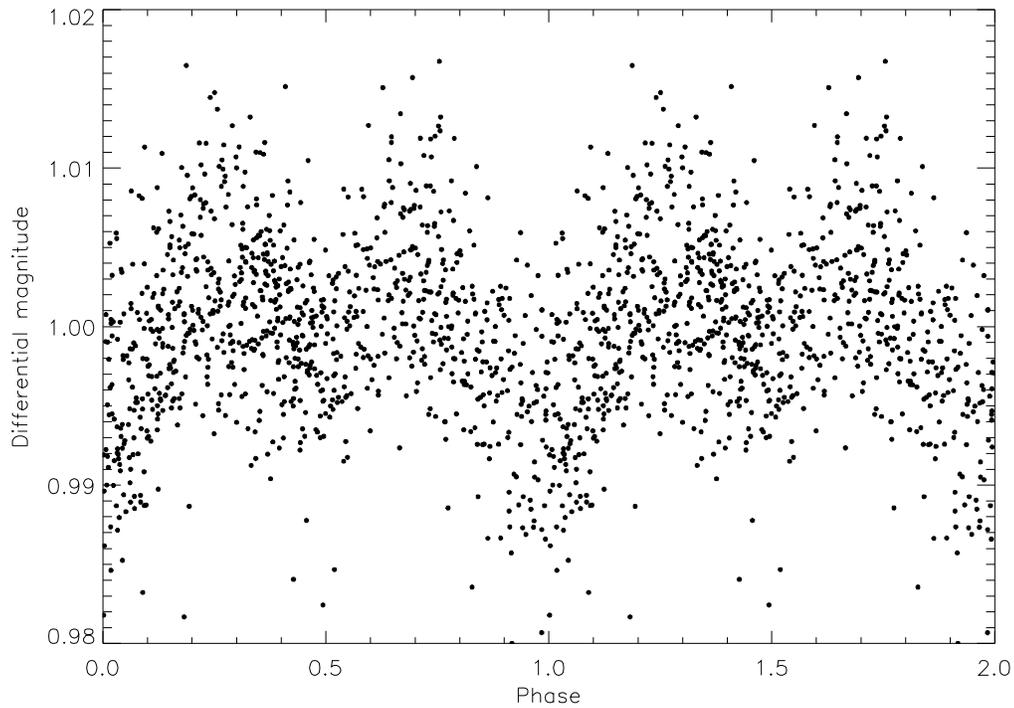

Fig. 14 - APACHE data for the star UCAC4 849-017521 , folded according to the period 0.5988 days.

**UCAC4 848-018678**

This star is in the same field of view of the previous variable, and for that we collected 2821 useful measurements during the same time span. Fig. 15 summarizes our results. By applying the GLS algorithm we found a peak periodicity at ~41 days. We folded the APACHE data to this period (Fig. 15), from which it appears that the light curve follows a sinusoidal-like modulation with ~0.02 peak-to-valley amplitude. This can be interpreted as a photometric variations related to the stellar rotation and due to an unevenly spotted photosphere. Due to our sparse measurements, nonetheless collected during a time span of nearly 15 months, we cannot be very confident that the true period is actually that we derived which is quite long and would require the monitoring of several consecutive rotations to be better constrained. In absence of any other information about this star, as mass, radius and age, we cannot use gyrochronology relations to support our interpretation. From the color indexes and color excess (see Tab. 1, and also J-H=0.701, H-Ks=0.223) we suggest that the star could be tentatively classified as an early-type M dwarf (Zombeck, 1990; Pecaut & Mamajeck, 2013). Photospheres of M dwarfs are generally expected to be spot covered, thus rotational modulations in the light curve are frequently detected, as we found for some of the APACHE targets. The UCAC4 catalog gives pmRA= -0.4±2.3 mas/yr and pmDE=1.4±2.6 mas/yr for the proper motion of this star, which is indeed not indicative of a nearby star. Therefore, another reliable possibility is that the star is actually a red giant, for which a small proper motion is highly probable. The color indexes also support this scenario, being compatible with a late KIII/early MIII star (Zombeck, 1990). If this is the case, this star could be assigned to the group of the semi-regular (SR) variables showing small-amplitude variations in its light curve.

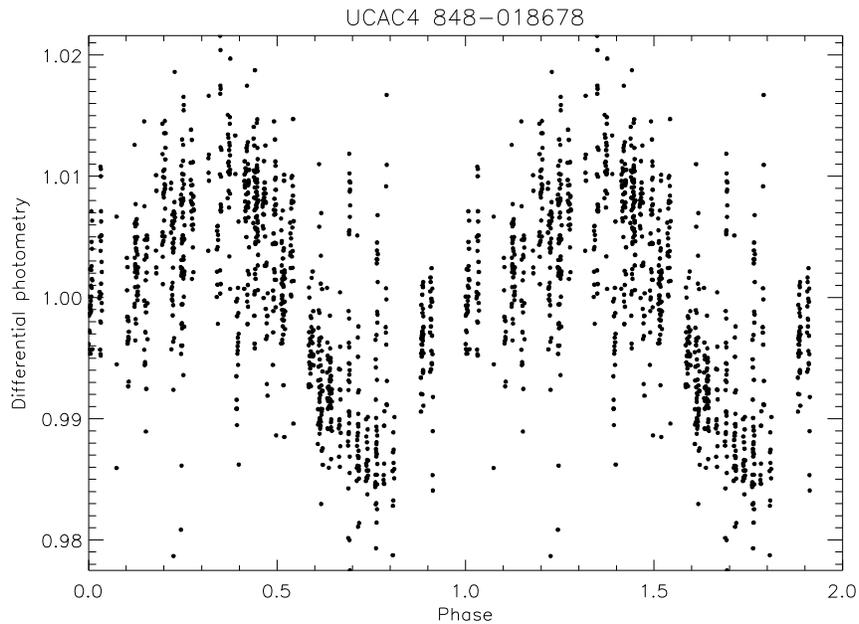

Fig. 15 - APACHE data for the star UCAC4 848-018678, folded according to the period P=41 days.

## UCAC4 849-017658

Belonging to the same field of the previous three variables, we collected for this object 2824 useful measurements in *I* band. By using the Lomb-Scargle algorithm we found half the real period to have the highest peak in the periodogram. By doubling this value and folding the light curve accordingly (P=0.35770 days), we recognized that its shape is clearly that of a short period eclipsing binary of EW type (Fig. 16). While the two maxima appear of the same height, taking into account the scatter in our data, the minima have slightly different depths (Δm~0.015 mag), as shown by the binned curve.

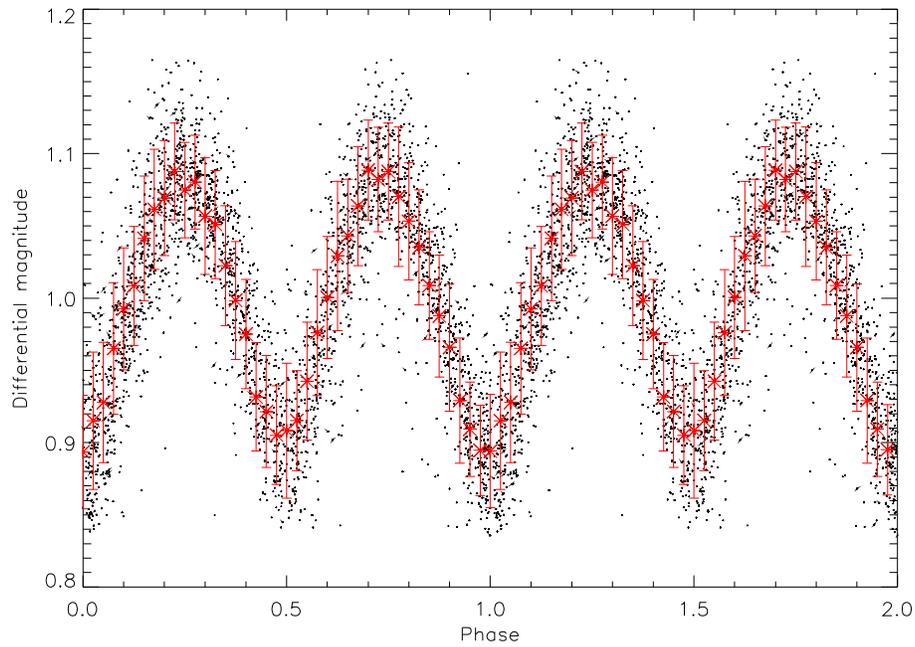

Fig. 16 - APACHE data for the star UCAC4 849-017658, folded according to the period 0.35770 days, with superposed the binned light curve in red (one bin corresponding to an interval of 0.025 in phase).

**Table 1** – Main information and results about the new variables discovered by the APACHE survey after the second season.

| Name | R.A.[a] | Dec.[a] | V | B-V | V-J | V-K | E(B-V)[b] | Period | Amplitude | T$_0$ | Var. Type | Gal. Latitude |
|---|---|---|---|---|---|---|---|---|---|---|---|---|
| | (J2000) | (J2000) | (mag) | (mag) | (mag) | (mag) | (mag) | (days) | (mag) | (HJD-2,455,000) | | (degrees) |
| UCAC4 837-000728 | 6.8337315 | +77.2397420 | 14.044±0.03 | 0.78±0.1 | 1.52±0.04 | 1.91±0.04 | 0.283 | 0.185848±0.000001 | ~0.2 (APACHE, *I* band) | 1367.2748975 | DSCT | +14.43° |
| UCAC4 612-044588  1SWASPJ092046.54+322044.1 | 140.1938968 | +32.3455925 | 13.196±0.03 | 0.68±0.04 | 1.33±0.04 | 1.85±0.04 | 0.017 | 0.88965±0.00001 | ~0.09 (APACHE, *I* band)  ~0.13 (SuperWASP) | 1300.683395 | EB[c] | +44.4° |
| UCAC4 667-058562 | 209.5897209 | +43.3495298 | 13.03±0.02 | 0.95±0.04 | 1.61±0.03 | 2.34±0.03 | 0.008 | 10.62±0.02 | ~0.07 (APACHE, *I* band)  ~0.026 (SuperWASP) | 1724.4068 | ROT | +68.8° |
| UCAC4 854-011628 | 270.4706342 | +80.6372137 | 15.13±0.01 | 0.70±0.04 | 1.37±0.03 | 1.76±0.05 | 0.076 | 0.30292±0.00001 | ~0.3 (APACHE, *V* band) | 1809.3644 | EW | +28.7° |
| UCAC4 609-091606 | 297.6272386 | +31.6943759 | 13.14±0.01 | 0.74±0.01 | 1.64±0.02 | 2.03±0.02 | 1.32 | 1.738±0.001 | >0.37 (APACHE, *I* band) | 1150.3817931 | EB | +2.6° |
| UCAC4 610-092815 | 297.6690953 | +31.9639217 | 15.74±0.01 | 0.78±0.10 | 1.51±0.03 | 1.95±0.05 | 1.10 | 0.42706±0.00001 | ~0.45 (APACHE, *I* band)  ~0.42 (APACHE, *V* band) | 1150.4765 | EW | +2.77° |
| UCAC4 621-119831 | 319.2999568 | 34.1319917 | 13.57±0.04 | 1.76±0.06 | 3.58±0.04 | 4.76±0.04 | 0.131 | - | >0.2 (APACHE, *V* band) | 1829.5151462 | L | -10.47° |
| UCAC4 620-119316 | 319.3728765 | +33.9577598 | 11.98±0.04 | 1.84±0.06 | 3.54±0.05 | 4.72±0.04 | 0.144 | - | >0.1 (APACHE, *V* band) | 1829.5151462 | L | -10.63° |
| UCAC4 620-119722  1SWASPJ211856.22+335439.3 | 319.7340386 | +33.9108506 | 14.11±0.11 | 0.43±0.11 | 0.96±0.11 | 1.25±0.11 | 0.15 | 0.2778±0.0001 | ~0.15 (APACHE, *V* band)  ~0.1 (SuperWASP) | 1385.653719 | DSCT | -10.88° |
| UCAC4 673-106048  1SWASPJ214537.33+442949.9 | 326.4054627 | +44.4972139 | 13.70±0.01 | 1.92±0.03 | 5.86±0.02 | 7.28±0.03 | 0.306 | - | ~0.2 (APACHE, *I* band) | 1532.6079559 | LB | -6.76° |
| UCAC4 858-013784 | 343.6031771 | +81.5279153 | 12.04±0.01 | 0.53±0.01 | 1.07±0.02 | 1.31±0.03 | 0.238 | 0.07077±0.00001 | ~0.02 (APACHE, *I* band) | 1385.653719 | DSCT | +19.67° |
| UCAC4 849-017521 | 348.9477371 | +79.6295775 | 12.18±0.03 | 0.47±0.03 | 1.08±0.04 | 1.40±0.04 | 0.198 | 0.5988±0.0001 | ~0.01 (APACHE, *I* band) | 1639.405 | EW | ~+17.6° |
| UCAC4 848-018678 | 349.0945677 | +79.4436050 | 13.06±0.05 | 1.54±0.08 | 2.80±0.05 | 3.72±0.05 | 0.2 | ~41 | ~0.02 (APACHE, *I* band) | 1367.2860 | ROT ? SR? | ~+17.6° |
| UCAC4 849-017658 | 350.5989427 | +79.7416248 | 14.36±0.09 | 0.80±0.08 | 1.40±0.08 | 1.840 | 0.187 | 0.35770±0.00001 | ~0.22 (APACHE, *I* band) | 1632.60 | EW | ~+17.6° |

**Notes.** The B and V magnitudes (and corresponding uncertainties) are taken from the UCAC4 catalog (Zacharias et al., 2013), while the J and K magnitudes (and corresponding uncertainties) come from the 2MASS survey (Skrutskie *et al.*, 2006).
The acronyms used for the classification of the variables follow the variable star type designations of the *AAVSO* International Variable Star Index (VSX) (see http://www.aavso.org/vsx/help/VariableStarTypeDesignationsInVSX.pdf).

*(a)* Celestial coordinates are taken from the UCAC4 catalog.
*(b)* Integrated along the line of sight.
*(c)* Possible secondary minimum visible in the APACHE data, with the folded light curve appearing distorted.

# Acknowledgments

This research has made use of several resources: the VizieR catalogue access tool and the SIMBAD database, operated at CDS, Strasbourg, France. We also used data from the first public release of the WASP data (Butters et al. 2010) as provided by the WASP consortium and services at the NASA Exoplanet Archive, which is operated by the California Institute of Technology, under contract with the National Aeronautics and Space Administration under the Exoplanet Exploration Program. Data from the NASA/ IPAC Infrared Science Archive were also used, which is operated by the Jet Propulsion Laboratory, California Institute of Technology, under contract with the National Aeronautics and Space Administration.
MD acknowledges partial support from INAF-OATo through the grant "Progetto GAPS: caratterizzazione spettroscopica e fotometrica dei target (attività cromosferica, rotazione) e studio delle sinergie tra GAPS e APACHE" (#35/2014), and from ASI under contract to INAF I/058/10/0 (Gaia Mission – The Italian Participation to DPAC). JMC and AB are supported by a grant of the European Union-European Social Fund, the Autonomous Region of the Aosta Valley and the Italian Ministry of Labour and Social Policy. We thank ASI also through contract I/037/08/0, and Fondazione CRT for their support to the APACHE Project.
The Astronomical Observatory of the Autonomous Region of the Aosta Valley is supported by the Regional Government of the Aosta Valley, the Town Municipality of Nus and the Mont Emilius Community.